\documentstyle[epsfig,aps,prl,twocolumn]{revtex}

\begin{document}

\tightenlines
\wideabs{

\title{Pressure-induced high-spin to low-spin transition in FeS\\
evidenced by x-ray emission spectroscopy}

\author{J.-P. Rueff, C.-C. Kao}
\address{National Synchrotron Light Source, Brookhaven National
Laboratory, Upton, NY 11973, USA}

\author{V. V. Struzhkin, J. Badro, J. Shu, R. J. Hemley, and H. K. Mao}
\address{Geophysical Laboratory and Center for High Pressure Research,
Carnegie Institution of Washington,\\
5251 Broad Branch Road NW, Washington, DC 20015, USA}

\maketitle

\begin{abstract}
We report the    observation  of the   pressure-induced   high-spin to
low-spin transition in  FeS using new  high-pressure synchrotron x-ray
emission spectroscopy techniques.  The  transition is evidenced by the
disappearance of the low-energy satellite  in the Fe K$\beta$ emission
spectrum of FeS.  Moreover,  the  phase transition is reversible   and
closely related to the  structural phase  transition from a  manganese
phosphide-like    phase to a   monoclinic  phase. The  study opens new
opportunities for investigating the electronic properties of materials
under pressure.
\end{abstract}

}
\narrowtext

The study of the  electronic structure of highly correlated transition
metal compounds has been an important subject in condensed-matter
physics over the last several decades.   The theoretical phase diagram
proposed by Zaanen, Sawatzky, and Allen \cite{Zaanen85}  is one of the
key steps  leading to  a  better understanding  of the materials.   In
addition to the on-site $d$-$d$ Coulomb  interaction ($U$) employed in
the original Mott-Hubbard theory, the ligand-valence band width ($W$),
the  ligand-to-metal  charge-transfer   energy  ($\Delta$), and    the
ligand-metal  hybridization interaction ($T$)  are explicitly included
as parameters in the model Hamiltonian. This classification scheme has
been  very  successful in  describing  the diverse properties and some
seemingly contradicting behavior of a large number of these compounds.
However, these   high-energy-scale charge fluctuations   are primarily
characteristic of the  elements  involved, and thus cannot  be  freely
adjusted for systematic study of  their effects, although they can  be
varied somewhat  by external temperature  and magnetic field.   On the
other hand, pressure can  introduce much larger perturbations of these
parameters than can either  temperature or magnetic field.  Hence,  it
is  of great interest  to  study the  high-pressure  behavior of these
systems, and specifically, to  correlate observed transformations with
changes in electronic structure.

Unfortunately,  high-pressure electronic  structure studies  have been
severely  handicapped  because   only  a   few  standard spectroscopic
techniques are compatible   with high-pressure cells.   Despite  rapid
developments in  synchrotron-based x-ray spectroscopy, there have been
limited investigations at high  pressure beyond a few x-ray absorption
studies   that have focused  mainly on  local  atomic coordination and
structure \cite{Itie}.   This   limited study  is  due  to  the strong
attenuation  of x-ray photons below 10  keV through  the pressure cell
(e.g., diamond  anvils),  the energy  range that  covers all the  3$d$
transition element K absorption edges, the  4$f$ rare earth L absorption
edges, and the 5$f$  actinides M  absorption edges,  as well as  their
characteristic  emission lines.  This problem can   be overcome by the
use   of  beryllium  as  a  gasket  material in   diamond  anvil cells
\cite{Hemley97},   and   allowing     both    the    incident      and
transmitted/scattered photons to pass through   the low absorbing   Be
gasket  instead  of  diamond. With this   technique, the  usable x-ray
energy   range can in principle  be  extended to include the important
spectral region from 3  to 10 keV,  making possible a wide  variety of
x-ray  spectroscopic studies of the  electronic structure of materials
under pressure.

Here we  present  a  study of the   high-spin  to low-spin (HS—LS) transition
in  FeS  using  new  high-pressure,  high-resolution  x-ray
emission spectroscopy (XES) techniques.   Under ambient  pressure, FeS
is     an             anti-ferromagnetic       insulator  (T$_N$=598K)
\cite{Horwood76,Shimada96}    and    has  a   NiAs-related  (troilite)
structure.   More   importantly, FeS falls    at  the boundary between
charge-transfer and  Mott-Hubbard insulators in  the ZSA phase diagram
($\Delta<U$ and $U$   relatively small \cite{Bocquet92}), so  that its
electronic structure is far   more   complicated, and has  been   very
controversial.   Pressure-induced structural phase  transitions in FeS
have been  extensively studied,  in    part because  the  material  is
considered to be a major component of the cores of terrestrial planets
\cite{Fei95,Sherman95,Kamimura92,King82,Kusaba97,Taylor70}.        FeS
undergoes  two structural  phase  transitions at ambient  temperature,
from the NiAs-related to a MnP-related structure  at 3.5 GPa, and then
to  a monoclinic phase at 6.5  GPa.   The latter transition is further
accompanied  by  an abrupt shortening  of the  $c$ parameter from 5.70
\AA\    to   5.54   \AA\  \cite{Fei95}.    Conventional  M\"{o}ssbauer
spectroscopy experiments \cite{King78,Kobayashi97} have shown that the
structural transition at 6.5  GPa is accompanied by the  disappearance
of the   magnetic  moment-induced  hyperfine  splitting.   Resistivity
measurements also hinted a possible insulator to semi-metal transition
at  3.5   GPa,  and perhaps   even   a  metallic  phase   above 7  GPa
\cite{Minomura63}.  We report  the observation of a HS--LS  transition
in FeS under pressure.  The spin  state of the Fe  in FeS is monitored
by high-resolution measurement of the  Fe K$\beta$ emission line.  The
emission spectrum of HS Fe is characterized by a main peak with energy
of 7058 eV, and a satellite peak located  about 12 eV lower in energy.
When the pressure exceeds  6.9 GPa, the satellite vanishes, indicating
the transition to the low spin-state with the  collapse of the Fe 3$d$
magnetic moment.

The experiments were carried out at the X25 hybrid wiggler beamline at
the National  Synchrotron  Light  Source (NSLS),  Brookhaven  National
Laboratory.    The beamline  consists of  a  Pt-coated double-focusing
mirror located upstream of a two-crystal monochromator.  Focused white
beam was used  to maximize the incident  x-ray flux for the excitation
of the  emission line.  The typical beam  size at  the sample position
was  about   0.5 mm$^2$.  The  spectrometer   for  the  x-ray emission
measurement  is a  1-meter Rowland  circle instrument  laid out in the
horizontal plane.  A spherically-bent Si(333)  single-crystal analyzer
crystal was used, with a Bragg angle corresponding  to the Fe K$\beta$
emission line (7058 eV) of 57.17$^\circ$. Use  of a 200 $\mu$m slit in
front of the detector produced an energy resolution of about 1.2 eV at
7 keV. Two-dimensional  scans  were carried  out to ensure  accurately
measured  lineshapes  \cite{Align}.   The  emission spectrum  was then
reconstructed by  extracting the maximum of  intensity in the detector
scan  at each  Bragg angle.  Typical  collection times   were about 90
minutes per spectrum.

High-purity   FeS was loaded into  a   diamond cell together with  4:1
mixture of methanol-ethanol as pressure  medium \cite{Synthesis}.  The
diamond-cell gasket was  initially a 5 mm  diameter  by 1 mm  thick Be
foil. The pressure was   measured by the ruby  fluorescence technique.
In order to reduce  background  signal, the  incident white  beam  was
collimated  by 150$\cdot$50~$\mu$m$^2$  water-cooled   1  mm  thick Ta
slits.  The slit assembly was in turn mounted on a water-cooled copper
block located in front  of the cell, which  was oriented vertically so
that  both  the incident and  collected x-rays  passed through  the Be
gasket.  The emitted  x-rays were detected  at a 90$^\circ$ scattering
angle  to further reduce   background signal.  To  evaluate the signal
from the Be gasket (known to contain small amount of Fe impurities), a
spectrum measured 150 $\mu$m from the sample  was subtracted from that
measured at the sample position.

\begin{figure}
\centerline{\epsfig{file=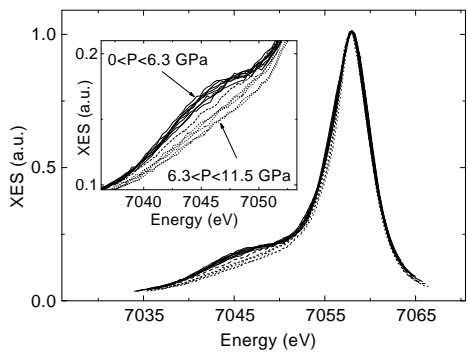,width=6cm}}
\caption{Fe  XES in FeS   as function of pressure.
The spectra were normalized with the intensity of the main peak set to
unity, and aligned  so that the (pressure independent)
energy of   the main peak is  set  to 7058 eV.    At low pressure, the
satellite at 7045.5 eV is characteristic of the high-spin state (solid
line) whereas  the absence of satellite  at  high pressure denotes the
transition  to  the low-spin state (dotted   lines).  The  dashed line
shows  an intermediate spectrum obtained  on decompression at 6.1 GPa.
The sequence of  pressures was  1.25, 2.0, 2.7,  3.5,  6.3, 7.5, 10.0,
11.5, 8.2 , 6.9, 6.1, 5.2, 1.4, and 0.2 GPa.}
\label{fesxes}
\end{figure}

Figure  \ref{fesxes} shows the Fe  K$\beta$  XES of
FeS between ambient pressure  and 11.5 GPa.   All spectra show  a main
peak  located at  7058 eV, usually  referred to  as the K$\beta_{1,3}$
line.  More importantly,  a well-defined  satellite located at  7045.5
eV,  denoted  K$\beta'$  in the following,    appears only in  spectra
measured for  pressures    below 6.3  GPa.   The  satellite  intensity
disappears in spectra for pressures  ranging between 6.3 and 11.5 GPa.
The width of  the main line  also shows significant narrowing  in this
pressure range.   The spectrum  taken at  6.1 GPa  shows  intermediate
behavior between these two  groups of spectra.   It is probably due to
the pressure gradient  in  the sample, which  is estimated  to be less
than 0.5 GPa.   Moreover, the observed  changes are  reversible as the
pressure reduced back to ambient.

\begin{figure}
\centerline{\epsfig{file=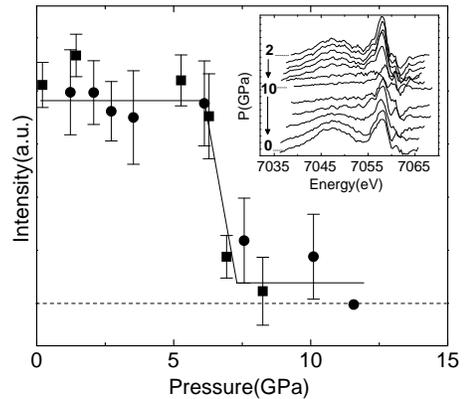,width=6cm}}
\caption{
Integrated intensity of the structure in the satellite region of the spectra shown in Fig.~1 obtained from a  Voigt profile fitted to  the difference spectrum upon compression (circles) and decompression (squares). The intensity was calculated from the difference spectra obtained by subtracting the  11.5 GPa spectrum from each scan (inset). The pressure in  the inset follows from top to  bottom the loop as indicated in the caption  to Fig.~1 and given on the left-hand side of the inset. The  second peak in 7056  eV region is  distorted by  the subtraction procedure; consequently  its intensity
is not shown. Spurious peaks in the K$\beta_{1,3}$ shoulder at 7058 to
7062 eV arise  from uncertainties associated with  the   subtraction
procedure.  The  solid line is  a guide  to the eye;  the dashed  line
shows zero intensity level.}
\label{loop}
\end{figure}

To illustrate the pressure-induced effects more clearly, the intensity
of the satellite  as a function of   the applied pressure is shown  in
Fig.  \ref{loop}.  The  intensity of the  7045.5 eV  satellite of each
spectrum was determined by subtracting  the spectrum at 11.5 GPa (with
no  satellite) and fitting  the    resulting profile with two    Voigt
functions (at 7045.5  and 7056 eV).   The difference spectra are shown
in the inset of Fig.~\ref{loop}.  In addition to the satellite peak in
the difference spectra,  a peak around  7056 eV is  observed, which is
also  a signature of  the HS state  (as discussed below). Although the
position  and intensity of  this peak are  subject to some uncertainty
due  to  the subtraction procedure,  the  peak disappears around 7 GPa
(like the 7045.5 eV   satellite).  From Fig. \ref{loop},  a reversible
transition occurs between 6.0 GPa and 7.0  GPa.  Since the uncertainty
in pressure due to the  relaxation processes and the pressure gradient
in the cell is about 0.5 GPa, the  observed change thus coincides with
the pressure of 6.5 GPa at which the magnetic splitting disappeared in
the M{\"o}ssbauer  measurement.   Notably,  there  is  no  appreciable
change  in     the satellite intensity   at   3.5  GPa,   the pressure
corresponding   to the   NiAs-   to MnP-    related structural   phase
transition.
 
To interpret  the observed spectral  changes,  a brief review  of  the
theory of  K$\beta$  x-ray emission spectra  is   given.  The K$\beta$
emission lines  are characteristic  x-ray  lines originating from  the
3$p$$\rightarrow$$1s$  decay.  For a  large number of transition metal
compounds, the K$\beta$  spectra  has  been interpreted  using  atomic
multiplet    calculations and configuration    interaction.  It is now
widely accepted that the spectral shape of  K$\beta$ emission line for
these  compounds is dominated by  final state  interaction between the
3$p$ core hole and  the electrons of  the partially filled 3$d$ shell.
Qualitatively, the main effect is    due to the exchange   interaction
between the core hole and the local moment, which results in splitting
of the  K$\beta$ spectrum into  HS  and LS final  states.  This simple
picture also predicts that the energy separation between the two peaks
is given  by the product  of  the exchange integral  $J$ and (2$S$+1),
where  $S$ is the  total spin of the 3d  shell; and that the intensity
ratio between the two is given by $S$/($S$+1) \cite{Tsutsumi76}.  Both
the   energy splitting    and   intensity ratio   are   modified  when
configuration interaction  is taken  into account \cite{Hermsmeier88}.
These        calculations          show                  that      the
$\underline{3p}^\uparrow$3$d^\uparrow$ final state is characterized by
a  single peak that  constitutes most   of  the intensity of the  main
emission        line.       On     the        other      hand,     the
$\underline{3p}^\downarrow$3$d^\uparrow$ final  state is further split
into two components, one  at significantly lower energy than predicted
by   simple  theory and one at   slightly  lower energy  than the main
emission line ($\underline{3p}^\uparrow$3$d^\uparrow$)  which  appears
as a  shoulder.   However, the simplified  picture does  point out the
qualitative changes  expected for the 3$d$ electrons  going from HS to
LS states, namely  smaller energy splitting between  the main peak and
the  satellite as well  as a reduction  in the  satellite to main peak
intensity ratio.

The  above theoretical approach has   been successfully applied to the
interpretation  of the  K$\beta$ emission  spectra  of a  number of Mn
\cite{deGroot95,Peng94a,Taguchi97},   and  Fe    \cite{Peng94b,Wang97}
compounds recently.  The  sensitivity of  the K$\beta_{1,3}$  emission
lineshape to  the spin state  of  the system is  demonstrated  Fig.~3,
where the Fe  K$\beta  _{1,3}$ line from   FeO and FeS$_2$ are  shown.
These spectra  were  recorded  at ambient  pressure,  and the  nominal
oxidation state is +2 for Fe in all three compounds; however, Fe is in
the HS  state   in FeO and  in   the LS state   in  FeS$_2$.  Note the
pronounced  satellite structure in  the K$\beta$  spectrum of FeO, and
the lack  of  a satellite  feature for  FeS$_2$.  Detailed  inspection
shows that there  is a small shoulder on  the low  energy side of  the
main peak for FeO,  indicating the presence  of the small HS component
discussed above.  On   the other hand, the  main  peak for FeS$_2$  is
narrower  and  more  symmetric, suggesting   that in  addition  to the
disappearance of  the   low energy satellite,  the    higher energy HS
component is absent as well.  These two examples clearly show that the
spectral shape  of  the  K$\beta$  emission line    can be   used   to
characterize the spin state    of the transition-metal ions  in  these
systems.

We can   now describe more  precisely the  pressure induced electronic
transition in  FeS.  Under ambient  condition, Fe in FeS is considered
to     be   divalent      with    a    HS     electron   configuration
($t_{2g\uparrow}^3e_{g\uparrow}^2t_{2g\downarrow}^1$).  This result is
now well  established   through both  M\"{o}ssbauer  spectroscopy and
magnetic susceptibility    measurements    \cite{Kobayashi97}.     The
M\"{o}ssbauer spectra clearly  indicate divalent Fe.  The isomer shift
of +0.76 mm/s  found  in FeS is  consistent with  Fe$^{2+}$ in the  HS
configuration,  which is further    confirmed by the large   effective
moment (estimated  to   be  5.5  $\mu_B$ by   magnetic  susceptibility
measurements     \cite{Horwood76}).     In     the     LS        state
($t_{2g\uparrow}^3t_{2g\downarrow}^3$),  the  two final  states of the
K$\beta$ emission  line,  $\underline{3p}^{\uparrow\downarrow}3d$, are
degenerate because of the absence of  a 3$d$ magnetic moment, thus the
XES spectrum reduces to a single peak as observed.

The result presented here is also consistent with the disappearance of
the  magnetic splitting  of the  hyperfine field  above 6.5 GPa  ({\it
cf.\/} Ref.\    \cite{Kobayashi97}).     It should     be  noted  that
M\"{o}ssbauer spectroscopy cannot unambiguously discriminate between a
local  diamagnetic or paramagnetic state.    The HS--LS transition may
also account for the shrinkage of the lattice cell  along the $c$ axis
associated  with  the  structural transition   observed around 7  GPa.
Indeed, it has been  proposed  that the spin-pairing   on the Fe  3$d$
shell might  cause a diminution of the   Fe radius \cite{Kusaba97} and
drives the observed transition. 

Metallization is a competing process that could also lead to the pressure-induced reduction of the satellite amplitude. However, metallization alone cannot explain the disappearance of the satellite above the transition pressure since a well defined satellite also appears in the K$\beta$ emission line of pure Fe-metal \cite{Holzer97} whose position and intensity is comparable to the one found in FeS at low pressure. Also, if metallization only were to occur, one would expect the satellite intensity of metallized high-spin FeS to be much stronger than that of metallic iron (the magnetic moments of high-spin FeS and metallic Fe are 5.5 and 2.2 $\mu_{B}$ respectively), which is not consistent with our measurements.

Whether  or not the    observed   transformation is  coupled    to   a metal-insulator
transition\cite{Resist} is still an open  question for both experiment and
theory.     Notably,  the  sequence   of transitions (e.g., HS--LS,
metallization, Mott transitions)  in such materials  generally is the subject of
much current  theoretical study\cite{Cohen97}.  

High-pressure measurements of the valence band emission and resonant inelastic x-ray
scattering are  in progress  to  clarify these issues.  X-ray emission
spectroscopy can also  be extended to measure symmetry-projected local
density  of states  of  the valence  band,   which provides  a  direct
measurement  of  the  anion   $p$-band  bandwidth   in  
transition-metal  compounds.   Pressure dependent measurements  of the
anion bandwidth  would  be important   in determining the   origin  of
metal-insulator transitions in these systems \cite{Chen93}.

We  thank N. Boctor for   sample  synthesis and characterization,  and
L. Berman, Z. Yin, and J. Hu for help with the experiments.  This work
was supported  by the NSF  and DOE (Contract  No. DE-AC02-76CH00016 to
NSLS).

\begin{figure}
\centerline{\epsfig{file=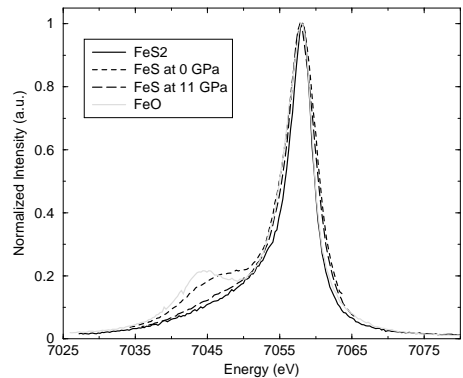,width=6cm}}
\caption{XES  spectra  of FeS   in  the high-spin  (room pressure) and
low-spin (11.5 GPa) states, along with room pressure reference spectra
of two  iron compounds with +2  oxidation states for  iron, namely FeO
(HS) and FeS$_2$ (LS).  The difference between the low-spin spectra of
FeS and FeS$_2$ is within the measured error.}
\label{fig3}
\end{figure}


\begin{references}
\bibitem{Align}  For the relatively low   Bragg  angles used in  these
measurements, the accurate  tracking   of analyzer  and detector   was
crucial  because  misalignment between the two   can cause an apparent
reduction in   intensity,  thereby modifying  the   measured  emission
lineshape.    For each Bragg angle,  the  detector was  scanned over a
range of 600 $\mu$m.
\bibitem{Synthesis}  Pure polycrystalline   FeS  was synthesized  from
high-purity  Fe  metal  and sulfur (both   99.999$^+$~\%) in evacuated
silica tubes.  It was annealed at 675~K for  several days until all of
the sulfur  reacted with the iron. The  tubes were then transferred to
1275~K  and annealed  for one  week  before  quenching rapidly in   an
ice-water bath.    The homogeneity   and  purity  of  the   sample was
confirmed by electron  microprobe analysis.  A  slight excess  of iron
was included to remain in the  troilite side of  the phase diagram and
to insure that the material is stoichiometric.
\bibitem{Resist} Although possible   metallization was  suggested   in
Ref.~14, only    a slight  increase  in resistivity    with increasing
temperature was measured.
\end{references}
\end{document}